\documentclass[epj]{webofc}
\usepackage[varg]{txfonts}   
\usepackage{hyperref}

\renewcommand*{\eqref}[1]{Eq.~(\ref{eq:#1})}

\newcommand*{\figref}[1]{Fig.~(\ref{fig:#1})}
\newcommand*{\figlab}[1]{\label{fig:#1}}

\newcommand*{\seclab}[1]{\label{sec:#1}}
\usepackage{color}
\wocname{ARENA-2016}
\woctitle{ARENA-2016}
\begin{document}
\title{Initial simulation study on high-precision radio measurements of the depth of shower maximum with SKA1-low }
%
%

\author{\firstname{Anne} \lastname{Zilles}\inst{1}\fnsep\thanks{\email{anne.zilles@kit.edu}}
\and \firstname{Stijn} \lastname{Buitink}\inst{2,3}
\and \firstname{Tim} \lastname{Huege}\inst{4}}


\institute{ Institut f\"u{}r Experimentelle Kernphysik, Karlsruher Institut f\"u{}r Technologie, 76128 Karlsruhe, Germany
\and
           Astrophysical Institute, Vrije Universiteit Brussel, Pleinlaan 2, 1050 Brussels, Belgium
\and
	Department of Astrophysics/IMAPP, Radboud University Nijmegen, 6500 GL, Nijmegen, The Netherlands
\and
 Institut f\"u{}r Kernphysik, Karlsruher Institut f\"u{}r Technologie, 76021 Karlsruhe, Germany.
          }

\abstract{%
As LOFAR has shown, using a dense array of radio antennas for detecting extensive air showers initiated by cosmic rays in the Earth's atmosphere makes it possible to
measure the depth of shower maximum for individual showers with a statistical uncertainty less than $20\,\mbox{g/cm}^2$. This allows detailed studies of the mass composition in the energy region around $10^{17}\,\mbox{eV}$ where the transition from a Galactic to an Extragalactic origin could occur.
Since SKA1-low will provide a much denser and very homogeneous antenna array with a large bandwidth of $50-350\,\mbox{MHz}$ it is expected to reach an uncertainty on the
$X_{\mbox{max}}$ reconstruction of less than $10\,\mbox{g/cm}^2$.
We present first results of a simulation study with focus on the potential to reconstruct the depth of shower maximum for individual showers to be measured with SKA1-low. In addition, possible influences of various parameters such as the numbers of antennas included in the analysis or the considered frequency bandwidth will be discussed.
}
\maketitle
\vspace{-0.5cm}
\section{Introduction}
\label{intro}
\vspace{-0.17cm}
With its about 70.000 antennas in a dense and uniform antenna spacing on a fiducial area of one km$^2$ and its large bandwidth of $50-350\,\mbox{MHz}$, the low-frequency part of
the the Square Kilometer Array (SKA1-low, first phase) will provide very precise measurements of individual cosmic ray air showers. An overview of the SKA1-low array as well as the physics motivation are described in~\cite{SKAoverview}. 
The LOFAR experiment shows that for individual shower events initiated by cosmic rays with energies of $10^{17}- 10^{17.5}\,\mbox{eV}$ the shower depth, which is correlated to the mass of the primary cosmic ray, can be reconstructed with a mean uncertainty of $16\,\mbox{g/cm}^2$~\cite{LofarNature}, for the ``best'' events even with $10\,\mbox{g/cm}^2$ uncertainty.
Based on results of first simulation studies, a reconstruction of the shower depth $X_{\mbox{\small max}}$ of individual showers with a mean intrinsic uncertainty of well below $10\,\mbox{g/cm}^2$ seems feasible.
\vspace{-0.3cm}
\subsection{Reconstruction method of the shower depth for SKA1-low}\seclab{subsec1.1}
\vspace{-0.3cm}
The presented method for the reconstruction of the shower depth of individual detected showers is adapted from the strategy developed for cosmic ray detection with the LOFAR experiment~\cite{LofarMethod} which is already successfully applied~\cite{LofarNature}. 

This simulation study on the possible uncertainty in the reconstruction of the shower depth by SKA1-low is based on sets of air-shower simulations induced by protons and iron nuclei. The air shower simulation was done by CORSIKA~\cite{CORSIKA}. The corresponding radio signal of the shower was calculated by CoREAS~\cite{CoREAS}, a plug-in of CORSIKA. 
As observer positions for which the radio signal was calculated 160 antenna positions in the shower plane defined by $\bf v\!\times\!\bf B$ and $\bf v\!\times\!(\bf v\!\times\!\bf B)$, the so-called star-shape pattern, were chosen, as shown in~\figref{fig-1} (left). Here, the vector $\bf v$ represents the velocity of the air shower front and $\bf B$ the Earth's magnetic field .
For each position the East-West and the North-South electric field components were filtered to the SKA1-low frequency band of $50-350\,\mbox{MHz}$. No antenna model was included. Afterwards, the time integral of the total received power was computed for each antenna in the star-shape pattern. 
The footprint of the radio signal with its asymmetry due to the interference of the Askaryan and the geomagnetic effect can be reconstructed by interpolation between the rays of the star-shape pattern (see~\figref{fig-1}, center). 
\begin{figure}[tb]
\centering
\includegraphics[width=0.325\textwidth,clip]{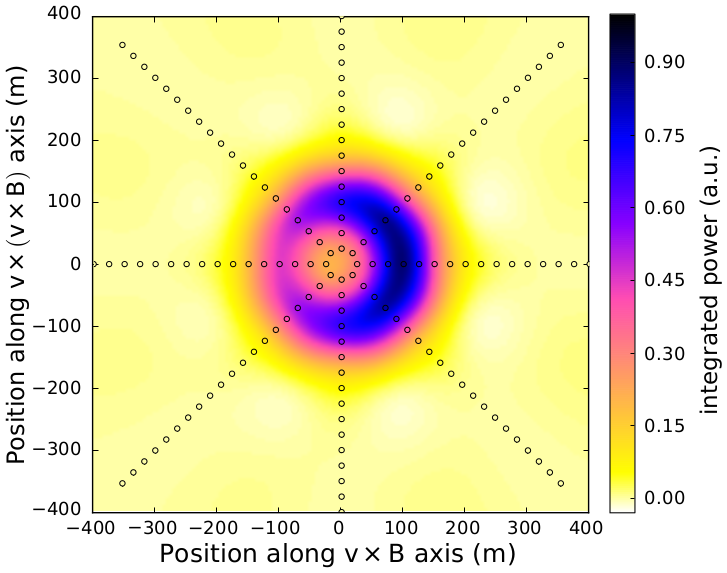}
\includegraphics[width=0.325\textwidth,clip]{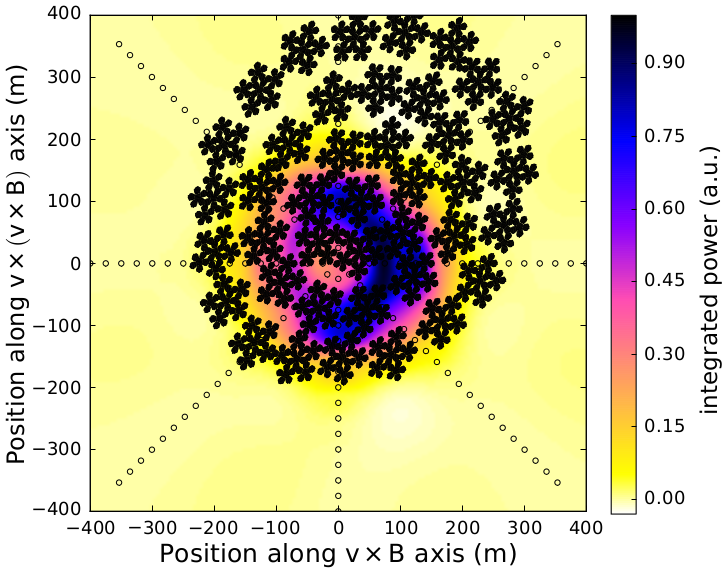}
\includegraphics[width=0.325\textwidth,clip]{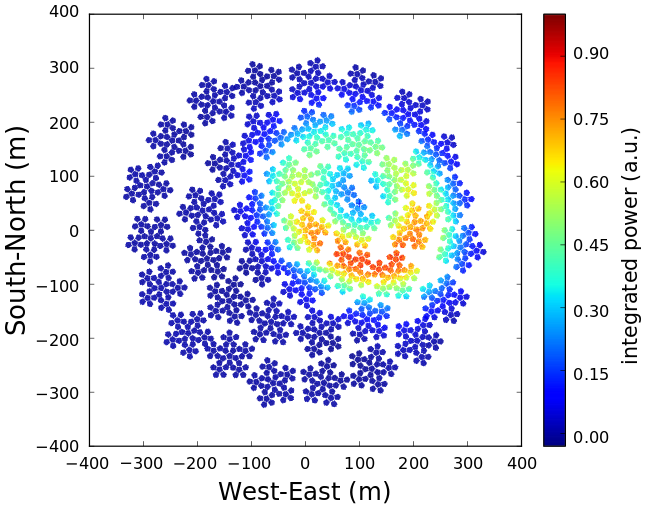}
\caption{The left and the center plots shows the total integrated power in the shower plane defined by the axes $\bf v\,\times\,\bf B$ and $\bf v\,\times\,(\bf v\,\times\,\bf B)$, while the right plot represents the integrated power on the ground with North pointing upwards.  Left: Total integrated power at the simulated antenna positions (circles) and the interpolated radio footprint in the shower plane for a proton-induced air shower with an primary energy of $E=10^{17}\,\mbox{eV}$ and a zenith angle of $36.87^\circ$, filtered to a frequency band of $50-350\,\mbox{MHz}$ and normalised to the maximum power. Center: The SKA1-low antenna positions rotated into the shower plane. At each antenna position, a value for the total received power is interpolated. Left: The example radio footprint on ground as detected by the SKA1-low array. The different color scheme is chosen to underline to change of the coordinate system.}
\figlab{fig-1}       
\end{figure}
To estimate the power which would be received by the SKA1-low antennas, the complete antenna array is rotated into the $\bf v\!\times\!\bf B$ and $\bf v\!\times\!(\bf v\!\times\!\bf B)$ frame and each antenna is assigned a total power corresponding to the interpolated power at this position (see ~\figref{fig-1}, right). These steps are performed for every simulation in a set which should consists in total of 50 proton-induced and 25 iron-induced air shower simulations.

Finally, each simulation acts once as ``fake'' data. This means that random noise is added to the interpolated total power of each antenna: 
5\% of the total power of each antenna as uncertainty on the antenna calibration and 1\% of the maximum power in the whole array as an approximation for Galactic noise.
This is just a first approximation since by this reconstruction the noise scales with the energy of the primary particle. More realistic assumptions have to be implemented in the future calculations.
The remaining simulations of the set are then compared to the ``fake'' data  on the basis of a  $\chi^2$ fit:
\begin{equation}
 \chi^2 =  \sum_{antennas} \left(\frac{P_{\mbox{\small fake}} - P_{\mbox{\small sim}}}{\sigma_{\mbox{\small fake}}}\right)^2
\end{equation}
Here, $P_{\mbox{\small fake}}$ represents the total power of a single antenna in the ``fake'' data set, with $\sigma_{\mbox{\small fake}}$ denoting the assumed uncertainty due to noise, and $P_{\mbox{\small sim}}$ the simulated power for the same position. This returns a $\chi^2$ value for every comparison of the ``fake'' data to one of the remaining simulations which can be connected to a corresponding $X_{\mbox{\small sim}}$ known from the simulation.

This leads to a distribution as shown in \figref{fig-2} (left) to which a parabola function can be fitted. This fit then returns the ``reconstructed'' shower depth $X_{\mbox{\small reco}}$ as the $X_{\mbox{\small max}}$ value for which the parabola has its minimum. The absolute difference between $X_{\mbox{\small reco}}$ and the each ``real'' shower depth $X_{\mbox{real}}$ of the simulated shower event which works as ``fake'' data  is filled into a histogram (see to \figref{fig-2}, right). In that histogram the $1\sigma$ uncertainty is defined at that value of $\mbox{abs}(X_{\mbox{\small reco}}-X_{\mbox{\small real}})$ which contains $68\%$ of entries. This returns the statistical uncertainty of the shower reconstruction shower by the method when adopting all simulations in a set are a ``fake'' event once.\\
\begin{figure}[tb]
\centering
\includegraphics[height=0.19\textheight,clip]{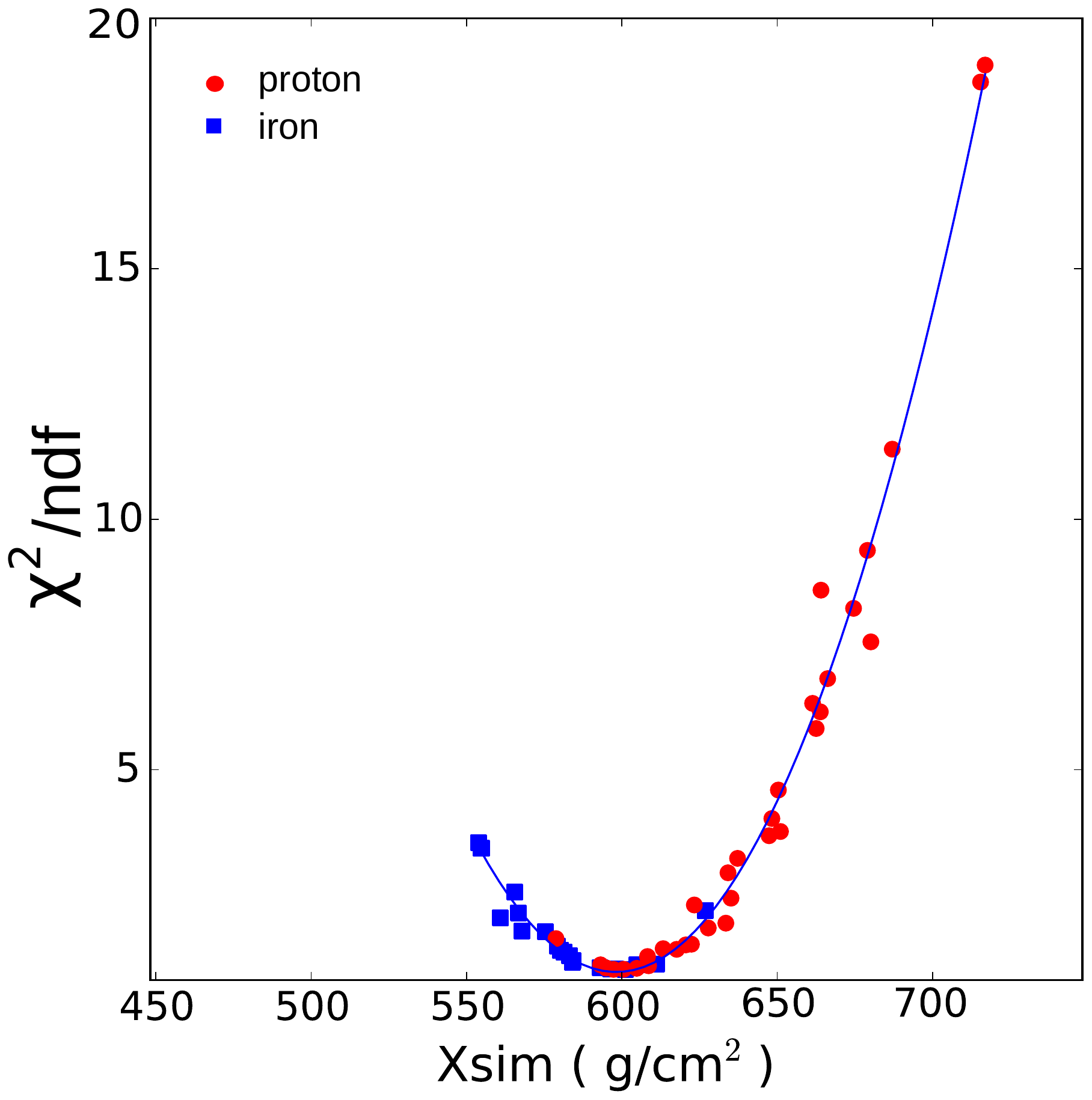}
\hspace{0.3cm}
\includegraphics[height=0.19\textheight,clip]{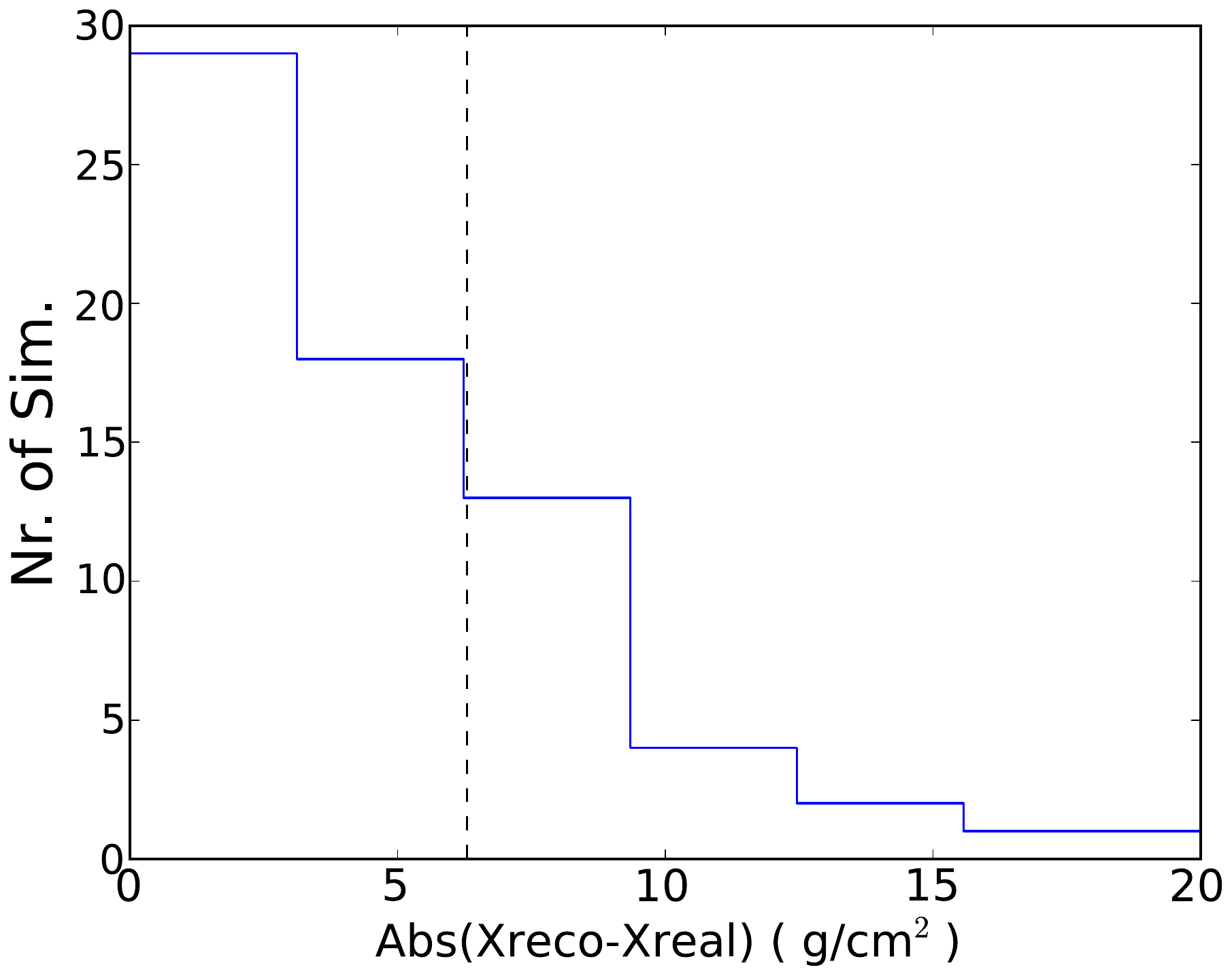}
\caption{Left: Example $\chi^2$ distribution for the SKA1-low array layout when comparing one ``fake'' event with the remaining simulations of the same set, filtered to $50-350\,\mbox{MHz}$. Right: Histogram for the differences of the reconstructed and the ``real'' shower depth: $\mbox{abs}(X_{\mbox{\small reco}}-X_{\mbox{\small real}})$. The $1\sigma$ uncertainty containing $68\%$ of the events returns a reconstruction uncertainty of $6.3\,\mbox{g/cm}^2$ for a shower of $E=10^{17}\,\mbox{eV}$ and  a zenith angle of $36.87^\circ$.}
\figlab{fig-2}       
\end{figure}
\vspace{-0.5cm}

\section{First results}\seclab{results}
\vspace{-0.17cm}
The results shown in \figref{fig-2} are based on $50\,$proton and $20$ iron showers with a primary energy of $E=10^{17}\,\mbox{eV}$, an azimuth angle of $225^\circ$ and a zenith angle of $36.87^\circ$. For these parameters the method returns an intrinsic reconstruction uncertainty of $6.3\,\mbox{g/cm}^2$.

Obviously, the uncertainty can depend on shower parameters such as the primary energy and the arrival direction.
For a fixed zenith angle but a higher primary energy of $E=10^{17.5}\,\mbox{eV}$, the simulation subset of $40\,$p and $18\,$Fe returns an uncertainty of $6.0\,\mbox{g/cm}^2$, a slightly smaller uncertainty. This could be explained by a possible better resolution of structures at the Cherenkov cone since the shower evolutions starts in average deeper in the atmosphere for higher primary energies. Then, this would be a benefit for the reconstruction method because in general the structures of the footprints of the ``fake'' and the simulated data get compared. This could mean that the more prominent the structures in the two dimensional profiles are the better the simulation which fits best can be fitted. 
For lower primary energies jitter in the $\chi^2$ distribution seems to get larger. This could be due to a flattening of the Cherenkov cone structure since the shower interacts earlier in the atmosphere. The example simulation subset of $40\,$p and $17\,$Fe for an energy of  $E=10^{16.5}\,\mbox{eV}$ leads to an intrinsic reconstruction uncertainty of $9.6\,\mbox{g/cm}^2$.
This is not an effect of the signal-to-noise ratio since this scales with the primary energy.

\begin{figure}[tb]
\centering
\includegraphics[height=0.19\textheight,clip]{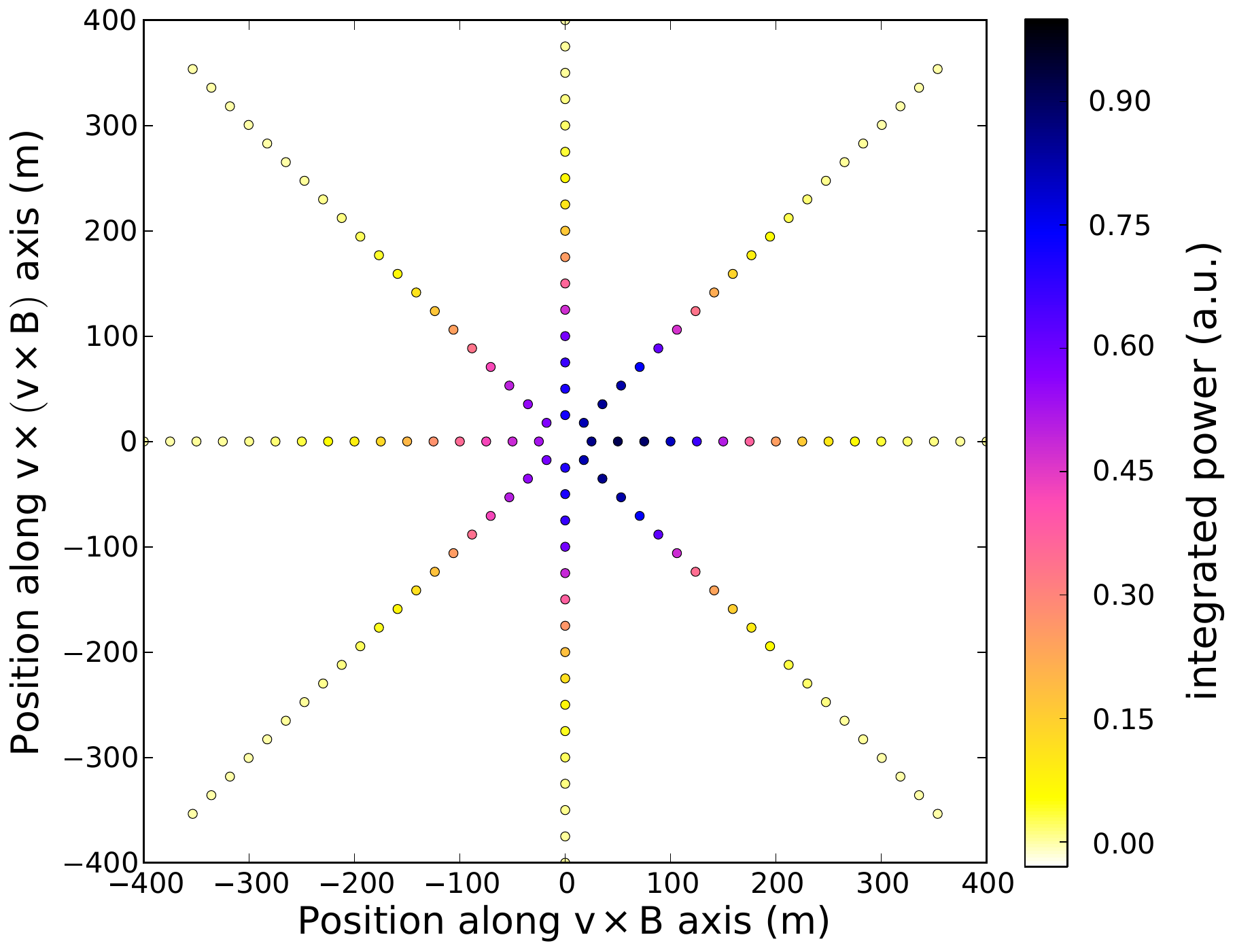}
\includegraphics[height=0.19\textheight,clip]{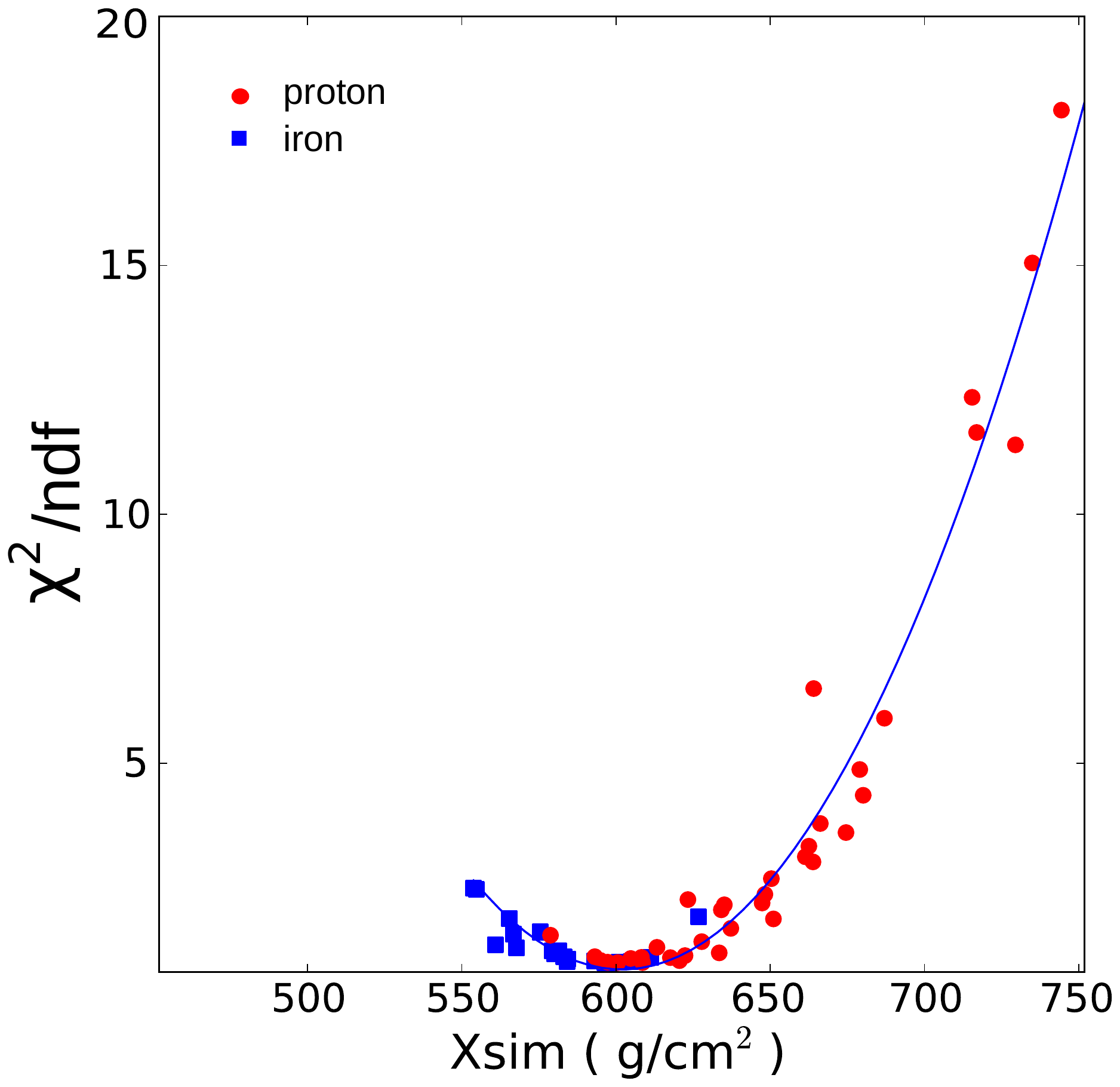}
\includegraphics[height=0.19\textheight, clip]{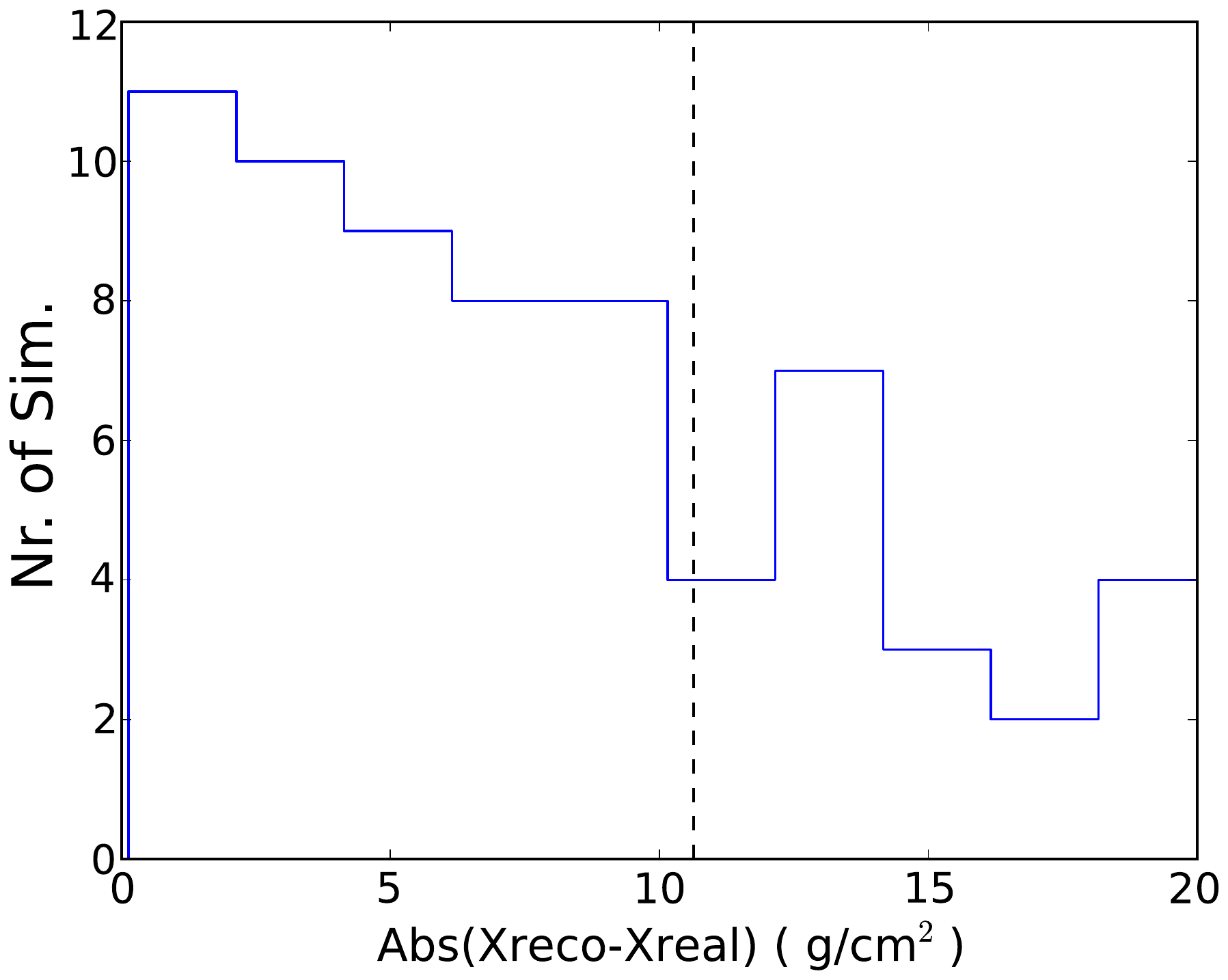}
\caption{Left: Radio footprint in $\bf v\!\times\!\bf B$ and $\bf v\!\times\!(\bf v\!\times\!\bf B)$ for a proton induced air shower with a primary energy of $E=10^{17}\,\mbox{eV}$ and a zenith angle of $36.87^\circ$, filtered to a frequency band of $30-80\,\mbox{MHz}$. Center: Example $\chi^2$ distribution for the SKA1-low array layout, filtered to the LOFAR bandwidth. Right: For these parameters the uncertainty of the shower depth reconstructions is $10.6\,\mbox{g/cm}^2$ ($1\sigma$).}
\figlab{fig-3}       
\end{figure}

One of the clear differences to the LOFAR experiment is the frequency band. In the LOFAR analysis the frequency band is limited to $30-80\,\mbox{MHz}$. In this frequency range, a bean-like structure is visible in the radio footprint instead of a clear Cherenkov ring as for the higher frequencies (see~\figref{fig-3}, left). This results in a slightly higher jitter around the parabola in the $\chi^2$ distribution as compared to the higher frequencies and finally in an intrinsic reconstruction uncertainty of $10.6\,\mbox{g/cm}^2$.

In addition, there are more aspects which influence the uncertainty of the shower depth reconstruction, like the numbers of antennas read out.
A thinning of the antenna array, meaning just reading out a subset of the antennas, does not influence the reconstruction precision as long as the read-out are homogeneously distributed over the radio footprint. The study of the impact of the number of the read-out antennas is closely connected to and therefore limited by the number of simulated antenna positions.

The method used for the power interpolation also turned out to be a limiting factor. As visible in \figref{fig-1} (left), dips in the interpolated power distribution are visible between the single rays. This seems to be an artifact of the interpolation since they vanish if more antennas on rays in-between the existing ones are simulated and included for the interpolation. 
\vspace{-0.1cm}
\section{Conclusion}
\vspace{-0.17cm}
The preliminary results of an initial simulation study show that a LOFAR-like approach for the reconstruction of the shower depth for single events is applicable to SKA1-low data. 
Since the simulation sets functioning as the basis for this study are still incomplete, the results can be interpreted as an indication of how precisely the SKA will reconstruct $X_{\mbox{\small max}}$. Hereby, the reconstruction profits especially from the very dense antenna array, but also from the larger bandwidth at higher frequencies in comparison to the LOFAR experiment. 
In this simulation study, we evaluated the intrinsic reconstruction accuracy reachable with a LOFAR-like analysis for data as it would be provided by SKA1-low. The SKA1-low array will achieve a mean intrinsic reconstruction uncertainty of less than $10\,\mbox{g/cm}^2$. The influence of experimental uncertainties (realistic background noise, uncertainty in the atmospheric properties, ...) on the reconstruction uncertainty has to be evaluated in a future study. 
Furthermore, it has to be studied whether the inclusion of signal timing, pulse shape and polarisation measurements would yield to additional information beyond the pure determination of $X_{\mbox{\small max}}$.

%
%
%

\end{document}